\documentclass[twocolumn,tighten]{aastex701}

\usepackage{amsmath}
\usepackage{newtxtext}
\usepackage{graphicx}
\usepackage[normalem]{ulem}

\usepackage[symbol]{footmisc}
\usepackage{multirow}
\usepackage{booktabs}
\usepackage{xcolor}

\def\arraystretch{1.19}

\RequirePackage{color}
\definecolor{cblue}{cmyk}{55,17,0,0}
\definecolor{darkgreen}{rgb}{0.0, 0.4, 0.0}
\definecolor{darkorange}{rgb}{0.5, 0.3, 0.0}
\definecolor{darkblue2}{rgb}{0.17, 0.49, 0.72}

\newcommand{\sunrise}{\mbox{\textsc{Sunrise\,iii}}}
\newcommand{\susi}{\mbox{\textsc{Sunrise\,iii}}/SUSI}

\usepackage{caption}
\DeclareCaptionFormat{cont}{#1 (cont.)#2#3\par}

\begin{document}
\correspondingauthor{Ivan Mili\'{c}}

\author[orcid=0000-0002-0189-5550,sname='Mili\'{c}']{Ivan~Mili\'{c}}
\affiliation{Institut für Sonnenphysik (KIS), Georges-Köhler-Allee 401a, 79110 Freiburg, Germany}
\affiliation{Faculty of Mathematics, University of Belgrade, Studentski Trg 12-16, 11000 Belgrade, Serbia}
\affiliation{Astronomical Observatory, Volgina 7, 11060 Belgrade, Serbia}
\email[show]{milic@leibniz-kis.de}

\author[orcid=0000-0002-2299-2800,sname='Osborne']{Christopher~M.~J.~Osborne}
\affiliation{SUPA School of Physics and Astronomy, University of Glasgow, G12 8QQ, UK}
\email{c.osborne.1@research.gla.ac.uk}

\author[orcid=0000-0003-1409-1145,sname='Iglesias']{Francisco~A.~Iglesias} \affiliation{Max-Planck-Institut für Sonnensystemforschung, Justus-von-Liebig-Weg 3, 37077 Göttingen, Germany}\affiliation{Grupo de Estudios en Heliofísica de Mendoza, CONICET, Universidad de Mendoza, Boulogne sur Mer 683, 5500 Mendoza, Argentina}\email{iglesias@mps.mpg.de}

\author[orcid=0000-0003-4319-2009,sname='Castellanos~Durán']{Juan~Sebastián~Castellanos~Durán}
\affiliation{Max-Planck-Institut für Sonnensystemforschung, Justus-von-Liebig-Weg 3, 37077 Göttingen, Germany}
\email{castellanos@mps.mpg.de}

\author[orcid=0000-0003-1670-5913,sname='Przybylski']{Damien~F.~Przybylski}
\affiliation{Max-Planck-Institut für Sonnensystemforschung, Justus-von-Liebig-Weg 3, 37077 Göttingen, Germany}
\email{przybylski@mps.mpg.de}

\author[orcid=0000-0002-9270-6785,sname='Chitta']{Lakshmi~Pradeep~Chitta}
\affiliation{Max-Planck-Institut für Sonnensystemforschung, Justus-von-Liebig-Weg 3, 37077 Göttingen, Germany}
\email{chitta@mps.mpg.de}

\author[orcid=0000-0002-0189-5550,sname='Chandra']{Sanghita~Chandra}
\affiliation{Max-Planck-Institut für Sonnensystemforschung, Justus-von-Liebig-Weg 3, 37077 Göttingen, Germany}
\email{cchandra@mps.mpg.de}

\author[orcid=0000-0003-1459-7074,sname='Lagg']{Andreas~Lagg} \affiliation{Max-Planck-Institut für Sonnensystemforschung, Justus-von-Liebig-Weg 3, 37077 Göttingen, Germany}\email{lagg@mps.mpg.de}	

\author[orcid=0000-0002-3418-8449,sname='Solanki']{Sami~K.~Solanki} \affiliation{Max-Planck-Institut für Sonnensystemforschung, Justus-von-Liebig-Weg 3, 37077 Göttingen, Germany}\email{solanki@mps.mpg.de}	

\author[orcid=0000-0001-6317-4380,sname='Riethmüller']{Tino~L.~Riethmüller}
\affiliation{Max-Planck-Institut für Sonnensystemforschung, Justus-von-Liebig-Weg 3, 37077 Göttingen, Germany}
\email{riethmueller@mps.mpg.de}

\author[orcid=0000-0002-9972-9840,sname='Gandorfer']{Achim~Gandorfer} \affiliation{Max-Planck-Institut für Sonnensystemforschung, Justus-von-Liebig-Weg 3, 37077 Göttingen, Germany}\email{gandorfer@mps.mpg.de}

\author[orcid=0009-0009-4425-599X,sname='Feller']{Alex~Feller} \affiliation{Max-Planck-Institut für Sonnensystemforschung, Justus-von-Liebig-Weg 3, 37077 Göttingen, Germany}\email{feller@mps.mpg.de}

\author[orcid=0000-0001-6029-7529,sname='Hoelken']{Johannes~Hoelken}
\affiliation{Max-Planck-Institut für Sonnensystemforschung, Justus-von-Liebig-Weg 3, 37077 Göttingen, Germany}
\email{hoelken@mps.mpg.de}

\author[orcid=0000-0003-3490-6532,sname='Smitha']{H.~N.~Smitha}
\affiliation{Max-Planck-Institut für Sonnensystemforschung, Justus-von-Liebig-Weg 3, 37077 Göttingen, Germany}
\email{narayanamurthy@mps.mpg.de}

\author[orcid=0000-0002-3387-026X,sname='del~Toro~Iniesta']{Jose~Carlos~del~Toro~Iniesta}
\affiliation{Instituto de Astrofísica de Andalucía, CSIC, Glorieta de la Astronomía s/n, 18008 Granada, Spain}
\affiliation{Spanish Space Solar Physics Consortium}
\email{jti@iaa.es}

\author[orcid=0000-0002-5054-8782,sname='Katsukawa']{Yukio~Katsukawa}
\affiliation{National Astronomical Observatory of Japan, 2-21-1 Osawa, Mitaka, Tokyo 181-8588, Japan}
\affiliation{Department of Earth and Planetary Science, The University of Tokyo, 7-3-1, Hongo, Bunkyo-ku, Tokyo 113-0033, Japan}
\affiliation{Department of Astronomical Science, The Graduate University for Advanced Studies (SOKENDAI), 2-21-1 Osawa, Mitaka, Tokyo 1818588, Japan}
\email{yukio.katsukawa@nao.ac.jp}

\author[orcid=0000-0002-0787-8954,sname='Bernasconi']{Pietro~Bernasconi}
\affiliation{Johns Hopkins University Applied Physics Laboratory, 11100 Johns Hopkins Road, Laurel, Maryland, USA}
\email{pietro.bernasconi@jhuapl.edu}

\author[sname='Berkefeld']{Thomas~Berkefeld}
\affiliation{Institut für Sonnenphysik (KIS), Georges-Köhler-Allee 401a, 79110 Freiburg, Germany}
\email{thomas.berkefeld@leibniz-kis.de}

\author[orcid=0000-0001-9228-3412,sname='Álvarez-Herrero']{Alberto~Álvarez-Herrero}
\affiliation{Instituto Nacional de T\'ecnica Aeroespacial (INTA), Ctra. de Ajalvir, km. 4, E-28850 Torrejón de Ardoz, Spain}
\affiliation{Spanish Space Solar Physics Consortium}
\email{alvareza@inta.es}

\author[orcid=0000-0001-5616-2808,sname='Kubo']{Masahito~Kubo}
\affiliation{National Astronomical Observatory of Japan, 2-21-1 Osawa, Mitaka, Tokyo 181-8588, Japan}
\email{masahito.kubo@nao.ac.jp}

\author[orcid=0000-0001-7764-6895,sname='Martínez~Pillet']{Valentín~Martínez~Pillet}
\affiliation{Instituto de Astrofísica de Canarias, Vía Láctea, s/n, E-38205 La Laguna, Spain}
\affiliation{Spanish Space Solar Physics Consortium}
\email{vmpillet@iac.es}

\author[orcid=0000-0001-8829-1938,sname='Orozco~Suárez']{David~Orozco~Suárez}
\affiliation{Instituto de Astrofísica de Andalucía, CSIC, Glorieta de la Astronomía s/n, 18008 Granada, Spain}
\affiliation{Spanish Space Solar Physics Consortium}
\email{orozco@iaa.es}

\author[sname='Grauf']{Bianca~Grauf}
\affiliation{Max-Planck-Institut für Sonnensystemforschung, Justus-von-Liebig-Weg 3, 37077 Göttingen, Germany}
\email{grauf@mps.mpg.de}

\author[sname='Carpenter']{Michael~Carpenter}
\affiliation{Johns Hopkins University Applied Physics Laboratory, 11100 Johns Hopkins Road, Laurel, Maryland, USA}
\email{michael.carpenter@jhuapl.edu}

\author[sname='Bell']{Alexander~Bell}
\affiliation{Institut für Sonnenphysik (KIS), Georges-Köhler-Allee 401a, 79110 Freiburg, Germany}
\email{albe@leibniz-kis.de}

\author[orcid=0000-0003-1483-4535,sname='Strecker']{Hanna~Strecker}
\affiliation{Instituto de Astrofísica de Andalucía, CSIC, Glorieta de la Astronomía s/n, 18008 Granada, Spain}
\affiliation{Spanish Space Solar Physics Consortium}
\email{streckerh@iaa.es}

\author[orcid=0000-0003-0175-6232,sname='Siu-Tapia']{Azaymi~L.~Siu-Tapia}
\affiliation{Instituto de Astrofísica de Andalucía, CSIC, Glorieta de la Astronomía s/n, 18008 Granada, Spain}
\affiliation{Spanish Space Solar Physics Consortium}
\email{siu@iaa.es}

\author[orcid=0000-0001-7094-518X,sname='Santamarina~Guerrero']{Pablo~Santamarina~Guerrero}
\affiliation{Instituto de Astrofísica de Andalucía, CSIC, Glorieta de la Astronomía s/n, 18008 Granada, Spain}
\affiliation{Spanish Space Solar Physics Consortium}
\email{psanta@iaa.es}

\author[orcid=0000-0002-2055-441X,sname='Blanco~Rodríguez']{Julian~Blanco~Rodríguez}
\affiliation{Universitat de Valencia Catedrático José Beltrán 2, E-46980 Paterna-Valencia, Spain}
\affiliation{Spanish Space Solar Physics Consortium}
\email{julian.blanco@uv.es}

\author[orcid=0000-0001-7452-0656,sname='Kawabata']{Yusuke~Kawabata}
\affiliation{National Astronomical Observatory of Japan, 2-21-1 Osawa, Mitaka, Tokyo 181-8588, Japan}
\email{kawabata.yusuke@nao.ac.jp}

\author[orcid=0000-0002-1043-9944,sname='Matsumoto']{Takuma~Matsumoto}
\affiliation{Centre for Integrated Data Science, Institute for Space-Earth Environmental Research, Nagoya University, Furocho, Chikusa-ku, Nagoya, Aichi 464-8601, Japan}
\email{takuma.matsumoto@gmail.com}

\author[orcid=0000-0002-4669-5376,sname='Ishikawa']{Ryohtaroh~T.~Ishikawa}
\affiliation{National Institute for Fusion Science, 322-6 Oroshi-cho, Toki City 509-5292, Japan}
\email{ishikawa.ryohtaro@nifs.ac.jp}

\author[orcid=0009-0002-6808-5154,sname='Harnes']{Edvarda~Harnes}
\affiliation{Max-Planck-Institut für Sonnensystemforschung, Justus-von-Liebig-Weg 3, 37077 Göttingen, Germany}
\email{harnes@mps.mpg.de}

\author[orcid=0000-0002-7044-6281,sname='Oba']{Takayoshi~Oba}
\affiliation{Max-Planck-Institut für Sonnensystemforschung, Justus-von-Liebig-Weg 3, 37077 Göttingen, Germany}
\affiliation{National Astronomical Observatory of Japan, 2-21-1 Osawa, Mitaka, Tokyo 181-8588, Japan}
\email{oba@mps.mpg.de}

\author[orcid=0000-0003-1971-5551,sname='Vukadinović']{Dušan~Vukadinović}
\affiliation{Max-Planck-Institut für Sonnensystemforschung, Justus-von-Liebig-Weg 3, 37077 Göttingen, Germany}
\email{vukadinovic@mps.mpg.de}

\suppressAffiliations

\title{\textsc{Sunrise iii}/SUSI reveals extended and spectrally structured Ca II\,K emission above the solar limb}

\begin{abstract}
The \susi{} instrument performed a sit-and-stare observation of the off-limb emission in the Ca~II~K line core, with a combination of excellent temporal and spectral resolution. The emission reaches up to 6~Mm above the solar limb, indicating the existence of off-limb structures akin to spicules. In this letter, we describe the spatio-spectral properties of the observed emission, focusing on the spectral line shape resolved in detail by \susi{}, and compare the observation with simple modeling that relies on a numerical simulation of the solar chromosphere. We find that the simulation contains enough cool and dense material at Mm heights to produce off-limb emission, but that the spectral width of these features is slightly lower than observed, necessitating more detailed modeling. 
\end{abstract}


\section{Introduction}\label{sec:intro}

The chromosphere is a layer of the solar atmosphere that connects the solar surface with the solar corona and acts as an intermediary for energy transport between the two. It hosts a plethora of dynamic small-scale phenomena termed spicules \citep{dePontieu_2004_Nature, Pereira_2016_spiculeshighres}, rapid blue/red excursions \citep{Kuridze_2015_rberre}, or fibrils \citep{dePontieu_2007_fibrilsSST, Kianfar_2020_fibrils}.  The common ingredient of spicules (observed off-limb) and rapid blue/red excursions (observed on-disk) is the injection of relatively cool and dense chromospheric material into coronal heights (a few Mm above the photosphere). The exact physical origin of spicules and their role in coronal heating is still a matter of debate. Many spicules display emission in hotter channels, suggesting they could serve as conduits for the upward transport of mass and energy \citep[e.g.][]{dePontieu_2009_heating, Henriques_2016_heating, Samanta_2019_heating}. Understanding the spatial and spectral properties of spicules and other small-scale chromospheric dynamics is, therefore, of high scientific importance.

Typically, these events are observed using narrow-band imaging \citep[e.g.][]{Sterling_2010_Hinode_spicules} or imaging spectrographs/spectropolarimeters with limited spectral resolution and spectral range \citep[e.g.][]{Pereira_2016_spiculeshighres}. Slit-based spectrograph observations from ground-based telescopes are difficult due to the limited image stability, low contrast, and lack of spectrum reconstruction for such observing geometry \citep[but see][]{Beck_2011_groundbased, De_pontieu_2012_CaIIHspicules}. The space-based spectroscopy of spicules is limited to the Mg~II~UV lines observed with the IRIS spectrograph \citep{IRIS_2014_IRIS}. The Sunrise Ultraviolet Spectropolarimeter and Imager \citep[\susi,][]{Feller_2025SoPh_SUSI, Iglesias_2025_SUSIcalib} onboard the \sunrise{} telescope \citep{Lagg_2025_SUNRISE3, Solanki_et_al_2026_thisspecialissue} provided a view of the lower solar atmosphere with exceptional spatial ($\approx0.1$~arcsec), and temporal ($0.25$~s) resolution, excellent spectral resolving power ($\approx 10^5$) and a large spectral range ($\approx 2~$nm). \sunrise{} builds on the earlier version of \textsc{Sunrise} \citep{bartholetal2011}, which flew twice: in 2009 \citep{solanki10} and in 2013 \citep{solanki17}.

\begin{figure*}
    \centering
    \includegraphics[width=0.45\textwidth]{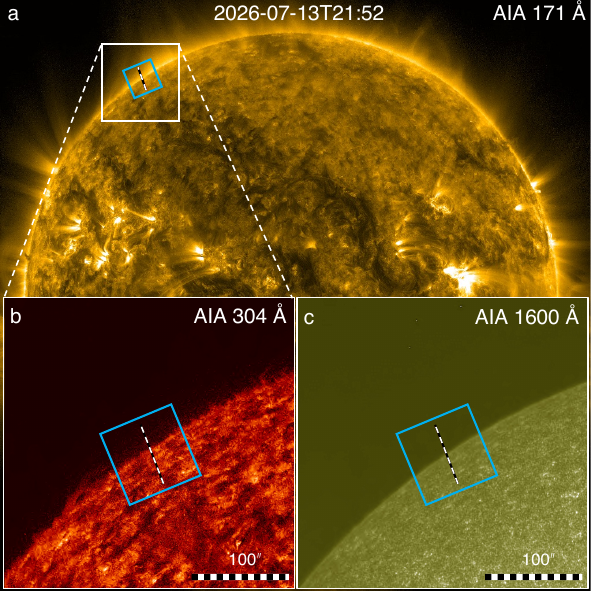}
    \includegraphics[width=0.52\textwidth]{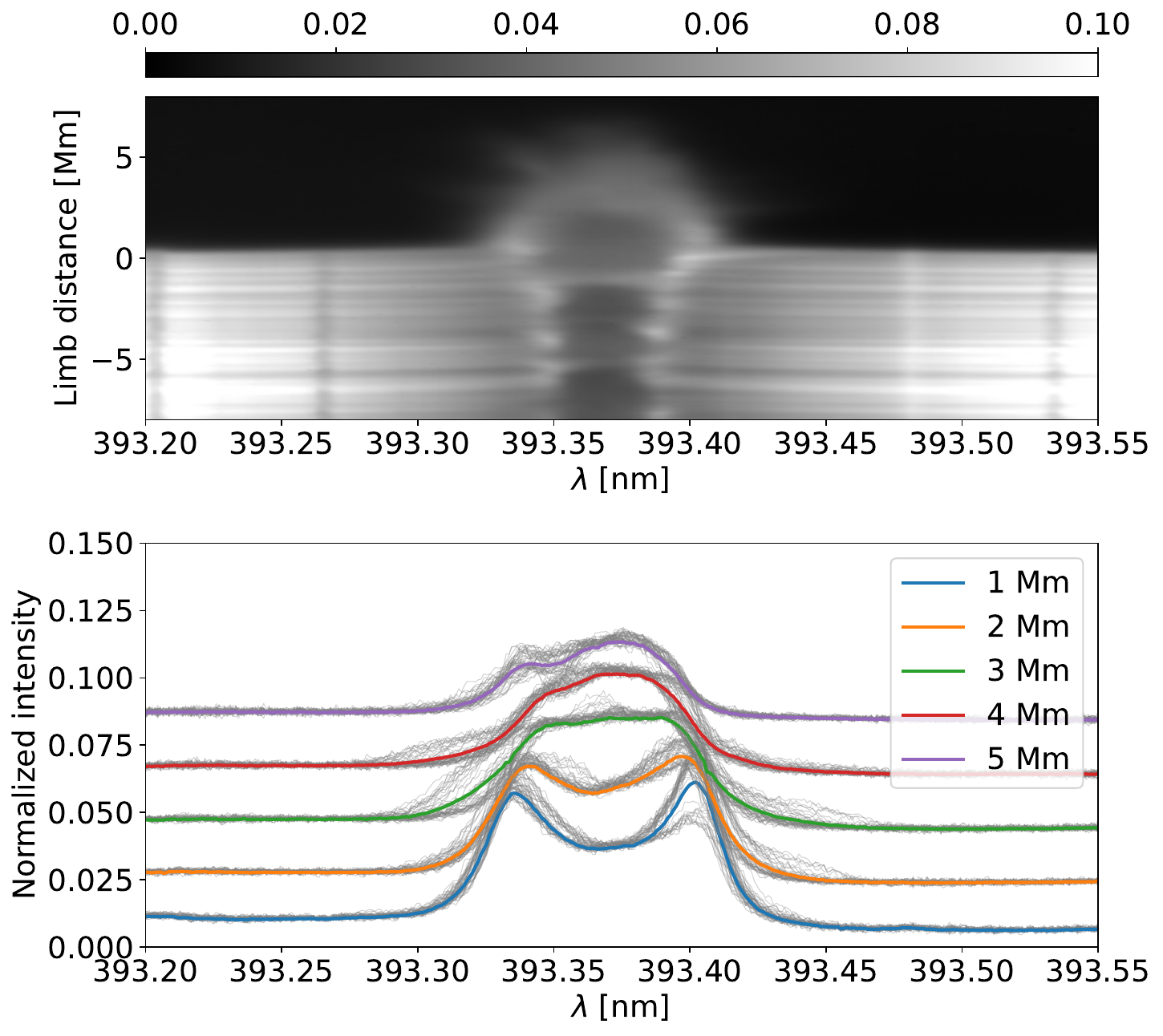}
    \caption{Summary of the observations. Left: AIA~17.1~nm, 30.4~nm, and 160.0~nm images, overlaid with a 60\arcsec{}$\times$60\arcsec{} field of view (blue square) and \susi{}\ slit position (dashed line). Top right: time-averaged Ca~II~K emission observed by \susi{}. Bottom right: Specific line shapes at different heights above the limb. Thick colored lines denote the time-averaged spectra, and thin gray lines outline the time-dependent spectra. Intensities are normalized to the quiet Sun at the disk center intensity at $\lambda=394.72~$nm and offset along the vertical axis for clarity. 
    }
    \label{fig:slit_and_line}
\end{figure*}

Here we report sit-and-stare observations in the wavelength domain around the Ca~II~K spectral line, often used to study spicules and related chromospheric phenomena, where the \susi\ spectrograph slit was oriented almost perpendicularly to the solar limb and captured the full extent of the line core emission. These observations reveal the rich spectral structure of the off-limb Ca~II~K emission, which serves as an indicator of the complex structuring of the solar chromosphere along the line of sight (LOS). This observing geometry results in a LOS that is roughly parallel to the solar surface and thus probes the horizontal distribution of plasma parameters. We contrast these spectral signatures with: i) the predictions of a canonical one-dimensional solar atmosphere model; and ii) the results of a simplified line formation synthesis that uses a recent radiative-magnetohydrodynamic (rMHD) simulation obtained with the chromospheric extension of the MURaM code \citep{Vogler_2005_muram, Przybylski_2022_muramche}. The main goal of this letter is to present the observations and compare the observed height and spectral width of the emission with a simplified but robust radiative transfer model applied to the simulations. Our results show that such a simulation produces a substantial amount of cool, dense plasma at heights required to produce Ca~II~K emission, but also underscore the necessity of a more detailed modeling approach to better understand and interpret chromospheric observations with very high resolution and cadence.

\section{Observations} \label{sec:obs}

The observations were taken from 21:53 UT to 21:55 UT on July 13th, 2024, with the field of view centered on $x=-544.0''$ (east limb) and $y=766.0''$ (northern hemisphere). The upper panel of Fig.~\ref{fig:slit_and_line} shows the SUSI field-of-view (FOV), and the slit position, overlaid on the co-temporal 17.1\,nm, 30.4\,nm, and 160.0\,nm images obtained by the Atmospheric Imaging Assembly \citep{Lemen_2012_AIA} onboard the Solar Dynamics Observatory \citep{Pesnell_2012_SDO}. The AIA images indicate that no large off-limb structures were present in the \sunrise{} FOV. The slit, oriented nearly perpendicular to the solar limb, was kept at a fixed position and recorded the spectrum in the wavelength region around the Ca~II~K line, spanning from 392.81 to 394.77~nm with 1~pm sampling. The temporal cadence of the observations is 0.25~s, which is an order-of-magnitude improvement over that of previous works. The top right panel of Fig.~\ref{fig:slit_and_line} shows the temporal average of the observed spectrum around the solar limb, which is defined as the inflection point of the continuum at the 394.72~nm wavelength. \sunrise{} conducted these observations from a 35~km altitude, resulting in only small amounts of stray light and an exceptionally clear and stable view of the solar limb \citep[see][for the details on the gondola, pointing, and wavefront correction]{Bernasconi_2025_sunrise, Berke_2026_sunrise}, comparable to the space-based observations conducted by Hinode/SOT \citep{Tsuneta_2008_Hinode_SOT}. 

The Ca~II~K emission is clearly visible above the limb, reaching up to and beyond 6 Mm heights (top right panel of Fig.~\ref{fig:slit_and_line}). These heights are substantially above the usual ranges for the Ca~II~K formation height in the canonical 1D models of the mean solar atmosphere \citep[e.g. FAL-C,][]{Fontenla_1993_falc}, which is assumed to be in the 1.5 to 2 Mm range. The bottom right panel of Fig~\ref{fig:slit_and_line} shows the change in the line shape when observing from the limb outwards. The emission aspect of the line exhibits two regimes. From the limb to $\approx2.5$~Mm above it, the line is an emission line with self-absorption in the line core. Beyond that, the line is a purely emission line with multiple peaks, which indicates several emission components along the line of sight. The height of the emission falls in the range previously reported for Ca~II spicules observed with Hinode/SOT \citep{Pereira_2012_quantifying_spicules} and SST/CRISP \citep{Scharmer_2003_SST, Scharmer_2008_CRISP} by \citet{Pereira_2016_spiculeshighres}. The observed line shapes and the emission height agree with the observations presented in \citet{Beck_2011_groundbased}.

\begin{figure}
    \centering
    \includegraphics[width=1.0\linewidth]{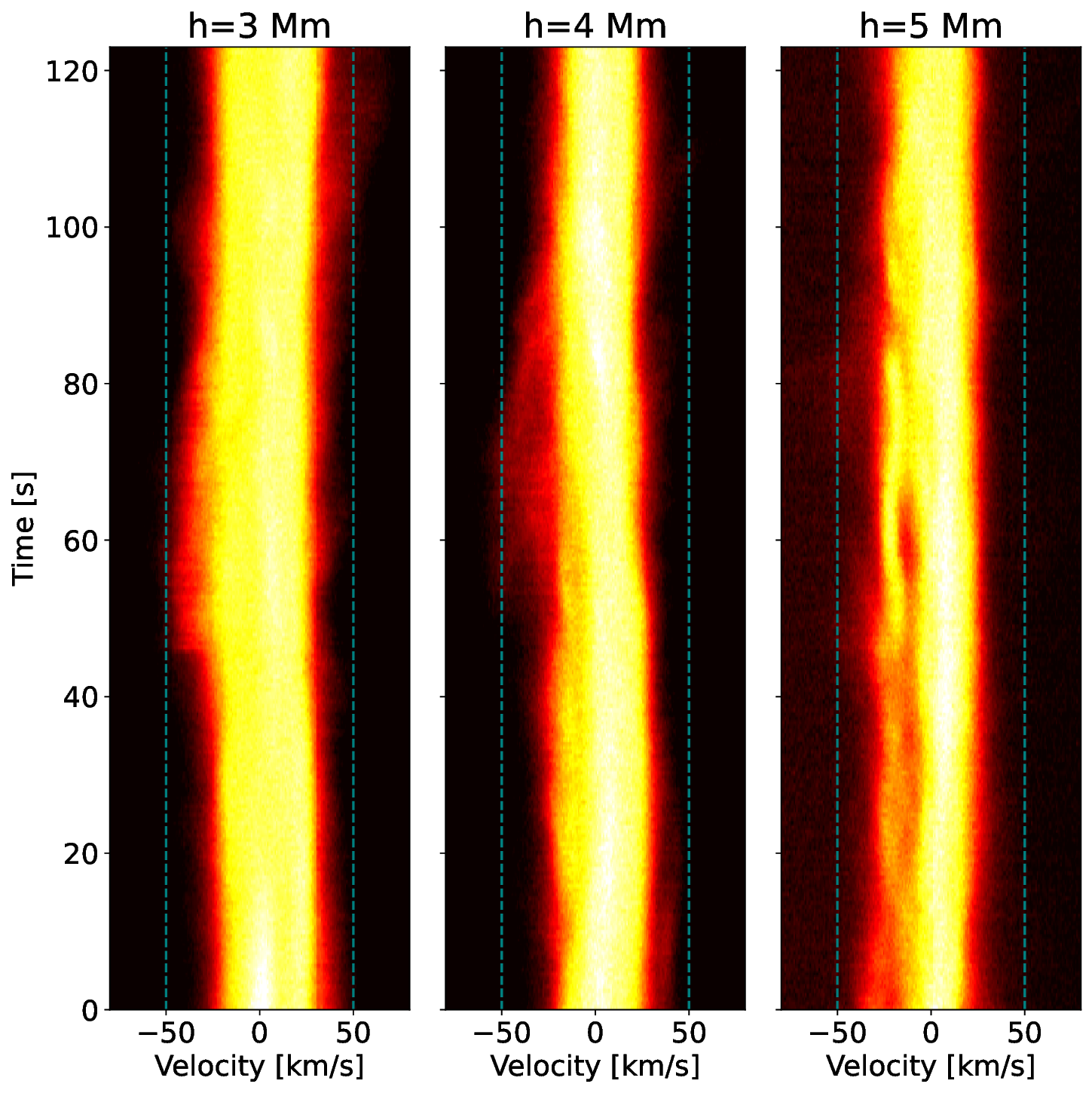}
    \caption{Time variation of the off-limb emission for different heights above the limb. Teal dashed vertical lines denote velocities of $\pm50$~km/s.}
    \label{fig:line_time_evolution}
\end{figure}

Although the observations cover a time span of only two minutes, the off-limb emission shows noticeable temporal variation (Fig.~\ref{fig:line_time_evolution}). This time dependence is likely due to different components entering and exiting the slit region. The width of the spectral line is largely unchanged, indicating a single, optically thick object or many optically thin emission features along the LOS. Fig.~\ref{fig:line_time_evolution} also shows weaker,  noticeably shifted emission components that reach LOS velocities of up to and above 50~km/s. These velocities are comparable to previously measured swaying and torsional velocities in type~II spicules \citep{De_pontieu_2012_CaIIHspicules}. Note that the limited duration of the observations prevents us from detecting periodicity in the line shifts.

In general, the emission above the limb arises when the LOS intersects a medium of substantial but finite optical thickness in the line, and negligible optical thickness in the continuum (such is the case, for example, in solar prominences). To model such an emission, a simple 1D plane-parallel model is inadequate \citep[e.g.][]{Milic_2012_spherical}, and we describe two simple modeling approaches below \citep[see also ][for a more statistical approach for modeling off-limb emission]{Tei_2020_IRIS}.

\section{Radiative Transfer modeling and comparison with the observations}

The Ca~II~K and H lines are resonance lines of singly-ionized calcium, which is the majority species of calcium in the lower solar atmosphere. This makes them one of the strongest spectral lines in the solar spectrum and a common diagnostic of solar activity and variability \citep{Ermolli_2018_CaIIK, Chatzistergor_2021_CaII_I}, or when observed in high-resolution, velocity and temperature diagnostics in the solar chromosphere, including flares  \citep[e.g.][]{Tamburri_2026_CaIIH}. They are formed in non-LTE \citep[][]{Mihalas_2014_book}, and are sensitive to the layers in the middle chromosphere where the gas density is low enough for three-dimensional radiative transfer effects to become important \citep{Bjorgen_2018_3D}. Such detailed modeling goes beyond the scope of this letter, and we resort to simpler approaches. 

\subsection{One-dimensional spherical atmosphere}

One way to reproduce off-limb emission in spectral lines is to employ a one-dimensional spherical model \citep[e.g.][]{Avrett_Loeser_1984_spherical, Kuridze_2022_tempmin}. Taking into account the sphericity of the solar atmosphere allows for the existence of lines of sight that intersect only a finite part of the atmosphere, thus properly reproducing the shape of the limb and the spectral line emission. We employed such a model by using the one-dimensional, plane-parallel FAL-C model of \citet{Fontenla_1993_falc} and the Lightweaver radiative transfer framework \citep{Osborne_2021_lw} to calculate the non-LTE level populations using the five-level Ca~II atom model of \citet{Shine_1974_CaII_model}, taking into account partial frequency redistribution \citep[PRD, ][]{Uitenbroek_2001_RH}. The depth- and wavelength-dependent opacity and emissivity coefficients are then integrated on a spherical grid, for the LOS directions corresponding to the \susi\ observations. The result is a synthetic spectrum that can be compared to the observed one. 

\begin{figure}
    \centering
    \includegraphics[width=1.0\linewidth]{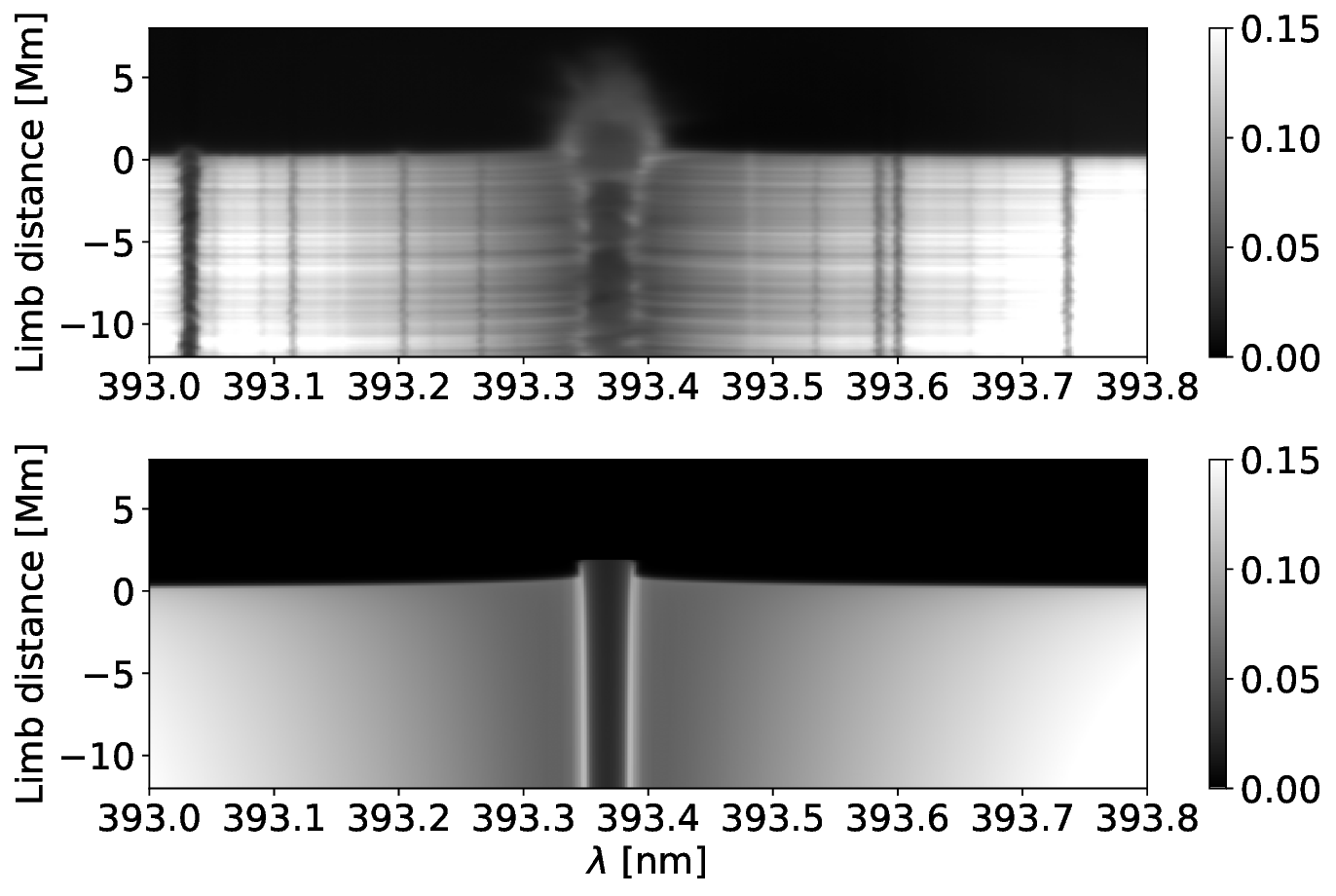}
    \caption{Top: Time-averaged Ca~II~K spectrum observed by \susi{}. Bottom: synthetic spectrum for the same spatial and wavelength domain, calculated from the FAL-C model using spherical geometry. 
    }
    \label{fig:1d_comparison}
\end{figure}

The comparison between the synthetic and observed spectra is given in Fig.~\ref{fig:1d_comparison}. The on-disk part of the spectrum shows good qualitative agreement, except for the pronounced broadening of the line at the extreme limb, which suggests that microturbulent velocity is either strongly height-dependent or anisotropic, as originally proposed by \citet{Holweger_1978_veldist}. Synthetic emission shows central self-reversal and extends approximately 1.5~Mm above the limb. However, there is no emission in the line above this point, which indicates that a mean one-dimensional solar model such as FAL-C is inadequate to appropriately describe the off-limb emission. This is because the FAL-C model contains a transition between the chromosphere and corona that occurs around $z=2~$Mm above the base of the photosphere. At coronal temperatures and densities, there is not enough Ca~II to cause substantial opacity in the K line, and therefore the line emission in the lower panel of Fig.~\ref{fig:1d_comparison} abruptly ends as the transition region begins. To remedy this problem and to understand the formation of this emission above the limb, we turn to a simulated three-dimensional atmosphere.

\subsection{Three-dimensional MURaM atmosphere}
\label{ssec:3dmodeling}

We consider two simulated snapshots 
obtained with the recent chromospheric extension  \citep{Przybylski_2022_muramche} of the MURaM code \citep{Vogler_2005_muram}. The models span 24~Mm$\,\times\,$24~Mm$\,\times\,$24~Mm, with a resolution of 23.46~km horizontally and 20~km vertically. The atmosphere ranges from -7~Mm below to 17~Mm above the photosphere, defined as the layer where the average optical depth in the continuum is unity. The simulation was initiated as a small-scale dynamo simulation and ran until stable \citep{Przybylski_2025_ssdch}. Subsequently, the simulation was modified by adding a bipolar magnetic field and further evolved. This magnetic configuration, referred to as the enhanced network, results in a large-scale magnetic loop and a small-scale structuring of the magnetic field in the lower layers. For more details on the enhanced network simulation run, see \citet{Przybylski_2022_muramche} and \citet{Ondratschek_2024_MgII}. This particular simulated atmosphere has been used to model the Mg~II~h\&k lines \citep{Ondratschek_2024_MgII,Ondratschek_2026_MgII} and Ca~II~854.2~nm lines as observed on-disk \citep{Ondratschek_2026_2026arXiv260202851O}. For this work, we consider an example snapshot from the small-scale dynamo (SSD) run and an example snapshot from the subsequent enhanced network (EN) simulation.

To model the Ca~II~K off-limb emission, we follow an approach similar to \citet{Chandra_2025_Halpha_spicules} \citep[see also][]{Chandra_2025_spicules_stat}. We first calculate the non-LTE populations of the Ca~II atoms in the 3D atmosphere using the Lightweaver framework, treating each column as a semi-infinite one-dimensional atmosphere (so-called 1.5D approximation). To mimic the off-limb view, we ignore the spatial variations of the source function in $x$ and $y$ and approximate the intensity using: 
\begin{equation}
    I_\lambda = \int_0^{s_{\rm los}} S_\lambda(s) e^{-t_\lambda} \chi_\lambda(s)ds,
    \label{eq:const_slab}
\end{equation}
where $t_\lambda = \int_0^s \chi_\lambda(s')ds'$ is the monochromatic optical depth along the line of sight, and $s_{\rm los}$ is the path the ray traverses along the line of sight, accounting for the curvature of the solar atmosphere. We obtain the source function $S_\lambda(s)$ by interpolating the $S_\lambda(z)$ calculated from the FALC model atmosphere under the PRD assumption, appropriately wavelength-shifted for local Doppler shifts. Given that FALC covers a range of approximately 0-2~Mm, we assume that the source function is constant with height above and below this region. This assumption essentially reduces the $S_\lambda$ to a single-scattering, optically thin approximation above 2~Mm heights. That is, we assume that the chromospheric material scatters the radiation and establishes a smooth source function variation, neglecting horizontal radiative transfer within the optically thick structures in the chromosphere. This way, we avoid dramatic variations of the source function among the nearby pixels, coming from the 1.5D approximation. As a consequence, we only partially model the self-absorption in the line core, which requires a gradient of the source function along the LOS. This gradient is still present, due to the sphericity of the solar atmosphere, similarly to the 1D case presented above.

\begin{figure}
    \centering
    \includegraphics[width=0.99\linewidth]{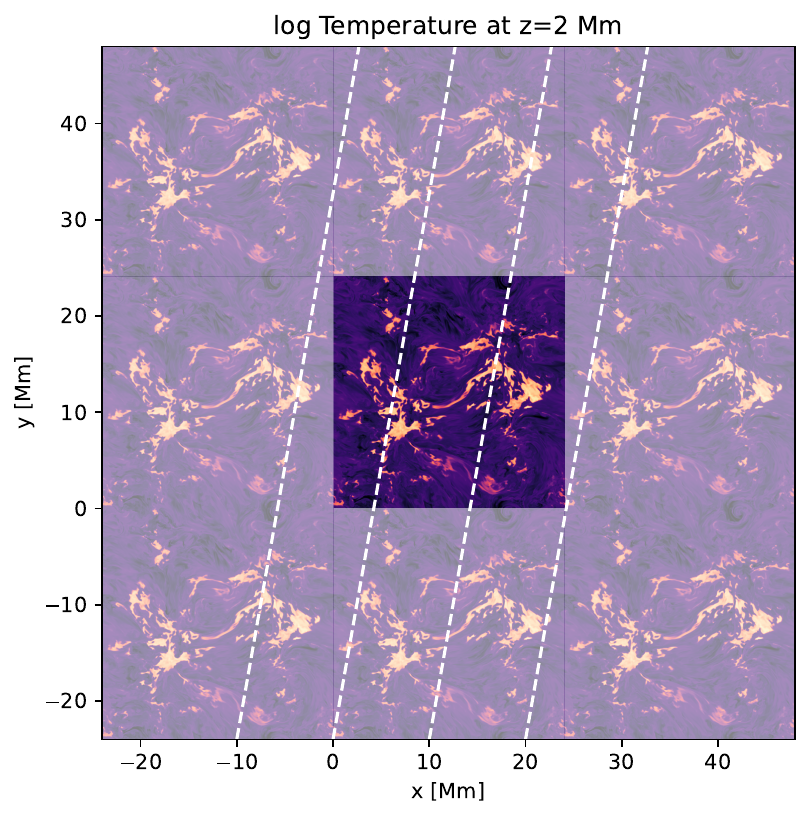}\\
    \includegraphics[width=0.99\linewidth]{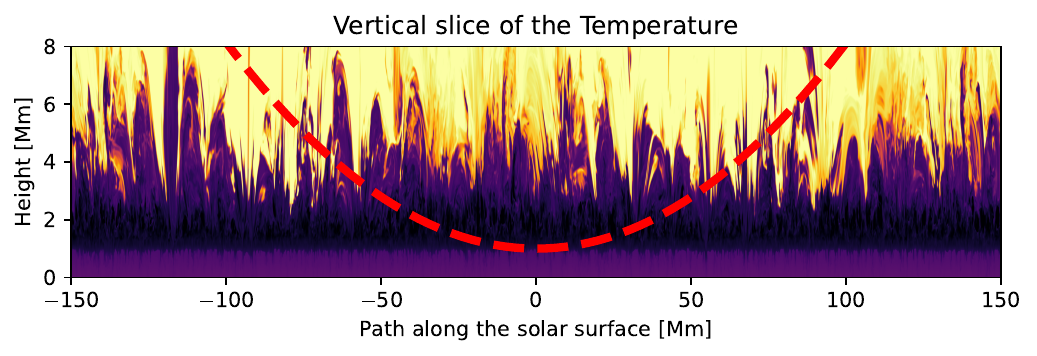}
    \caption{Line-of-sight path calculation for off-limb modeling using the 3D atmospheric model. Top: horizontal stacking of the cube and several example lines of sight. Only $3\times3$ cubes are plotted, for clarity. Bottom: temperature stratification in an example vertical slice with an example curved line-of-sight plotted as a red dashed line. The plot is compressed along the horizontal axis for clarity.}
    \label{fig:los_extraction}
\end{figure}

The optical depth is obtained by integrating the spectral line opacity:
\begin{equation}
    \chi_\lambda^L = n_l B_{lu} \frac{h\nu}{4\pi} \phi_\lambda,
    \label{eq:opacity}
\end{equation}
along the LOS and adding the continuum opacity due to neutral and negative hydrogen ions \citep{Grey_2005_book}. Here $n_l$ is the population of the lower level, $B_{lu}$ is the Einstein coefficient of absorption, and $\phi_\lambda$ is the line absorption profile. The line absorption profile is assumed to have a Voigt shape, which includes thermal broadening, radiative and van der Waals damping,  and Doppler shifts due to motions along the LOS, taken directly from the simulation. The emergent intensity is then obtained through the following steps:
\begin{itemize}
    \item The simulated cube, including the calculated level populations, is periodically stacked in the horizontal direction to obtain an atmosphere that is effectively infinite in $x$ and $y$, with $z$ as the atmospheric normal.
    \item Vertical 2D slices are extracted at a slanted angle of 10 degrees with respect to the $y$ axis, to minimize periodicity of resulting thermodynamic structures.
    \item Curved rays, corresponding to the curvature of the solar atmosphere at the limb, are extracted, and the opacity given by Eq.~\ref{eq:opacity} is integrated to obtain the total optical path along the LOS. The source function is interpolated onto the ray, with constant extrapolation where necessary.
    \item Finally, the emergent intensity is calculated for each LOS using Eq.~\ref{eq:const_slab}. This calculation, for a single vertical slice, results in a single synthetic slit.
\end{itemize}
This process is summarized in Fig.~\ref{fig:los_extraction}. For a cube that extends $\approx$ 8~Mm in height, the effective length of the line of sight at the limb is of the order of 200~Mm. This is roughly two orders of magnitude longer than the typical range probed by on-disk observations.

\section{Results and Discussion}
\label{sec:results}

\begin{figure}
    \centering
    \includegraphics[width=0.99\linewidth]{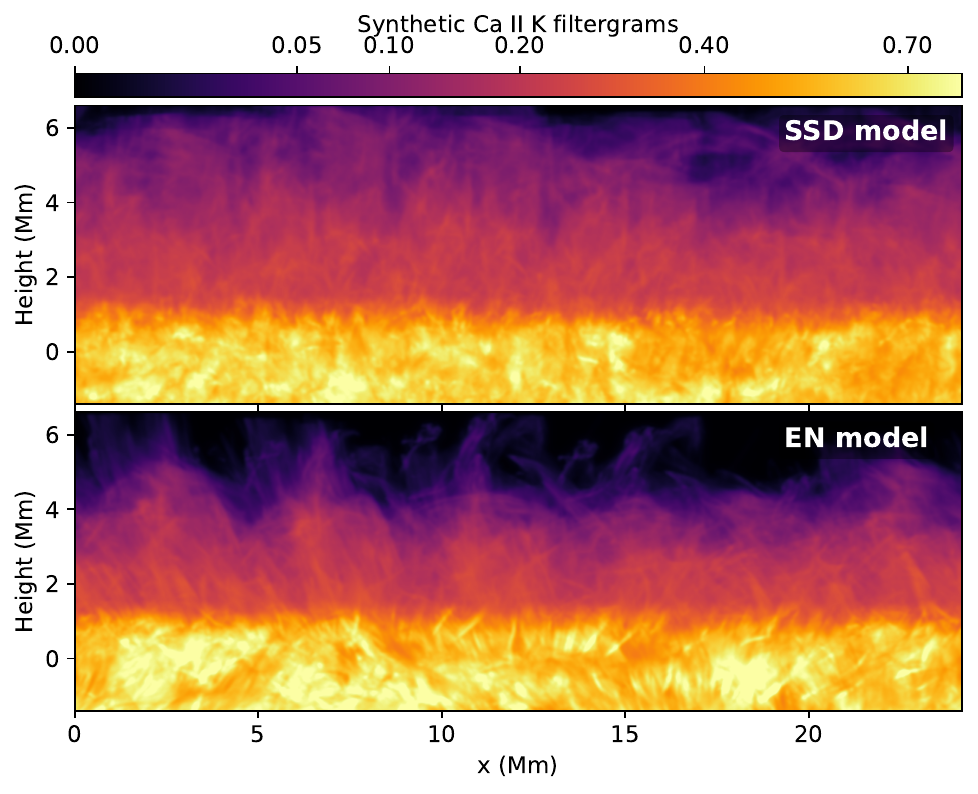}
    \caption{Synthetic Ca~II~K filtergrams obtained from the two considered simulated atmospheres. Top: small-scale dynamo simulation. Bottom: enhanced network simulation. The color scale is non-linear to better show off-limb emission.}
    \label{fig:filtergrams}
\end{figure}

The synthetic slit observations were calculated for each $x-$ position in the cube, following the procedure outlined in Sec.~\ref{ssec:3dmodeling}, resulting in a three-dimensional datacube $(x,h,\lambda)$ where $h$ is the height relative to the limb. The data have then been spatially convolved with the Airy point spread function of a 1-m telescope and spectrally convolved with a Gaussian kernel corresponding to the \susi{} spectral resolution. Fig.~\ref{fig:filtergrams} shows synthetic filtergrams obtained by spectrally integrating the emergent intensity over a wavelength range of $\approx0.1$~nm around the line core. The images show good qualitative agreement with the CaII~H observations using Hinode/SOT \citep{Suematsu_2008_hinodespicules, Pereira_2014_spicules} and  SST/CRISP \citep{Pereira_2016_spiculeshighres}. The most evident result is that the chromospheric emission from the SSD run reaches around 1.5~Mm higher on average than in the EN run. This is presumably because the stronger magnetic field found in the EN atmosphere becomes horizontal relatively low in the atmosphere, preventing the cool, dense plasma necessary to produce the Ca~II emission from reaching larger heights. On the other hand, the filtergram from the EN run shows more pronounced spicular structures. 

\begin{figure*}
    \centering
    \includegraphics[width=0.99\textwidth]{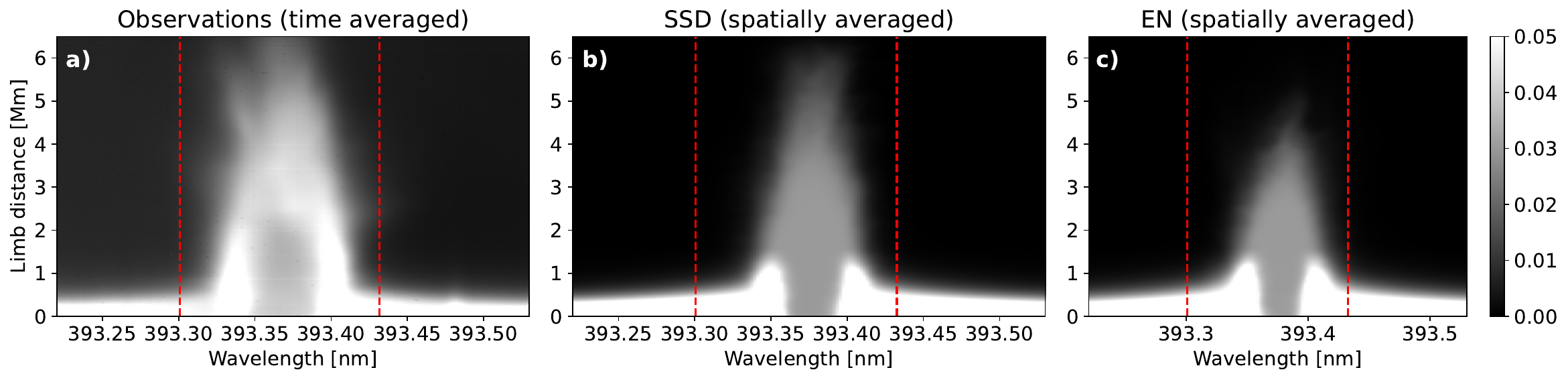}
    \caption{Comparison between the observed and simulated off-limb emission. Panel a): Temporally averaged emission above the limb observed by \susi{}. Panels b) and c): Spatially averaged synthetic slits for the SSD and the EN simulations. Vertical red dashed lines indicate wavelengths where the velocity corresponds to $\pm50$~km/s.}
    \label{fig:spectra_comparison}
\end{figure*}

Fig.~\ref{fig:spectra_comparison} shows the comparison of the temporally averaged observed spectra and synthetic spectra calculated from the SSD and EN snapshots and averaged along the $x-$axis. While such averaging does not represent a one-to-one correspondence between the observations and simulations, it serves as a first comparison of the two. The vertical extent of the emission in the simulated spectra is much higher than in the 1D case, indicating the MURaM atmospheres contain a substantial amount of cool and dense plasma at 2-6~Mm heights. The presence of this material can be explained by spicule-like structures already found in the chromospheric MURaM simulations by \citet{Chandra_2025_Halpha_spicules,Chandra_2025_spicules_stat}. The spectrum synthesized from the SSD run shows an emission height comparable to the observed one, but the line appears narrower. In the EN case, both the height of the emission and the line width are lower than in the observations. Both synthetic spectra exhibit weaker intensity of the emission, which can be attributed to our simplified treatment of the source function. Specifically, the source function is likely to increase if the radiative transfer within the spicules is taken into account. The same effect can explain the lower height of the self-absorption peaks in the synthetic spectra. Finally, large velocities of the chromospheric material would cause Doppler brightening (an increase in the amount of absorbed and re-emitted intensity due to large Doppler shifts), and thus the source function. Such modeling requires self-consistent 3D radiative transfer, which is beyond the scope of this letter.

\begin{figure*}
    \centering
    \includegraphics[width=0.99\textwidth]{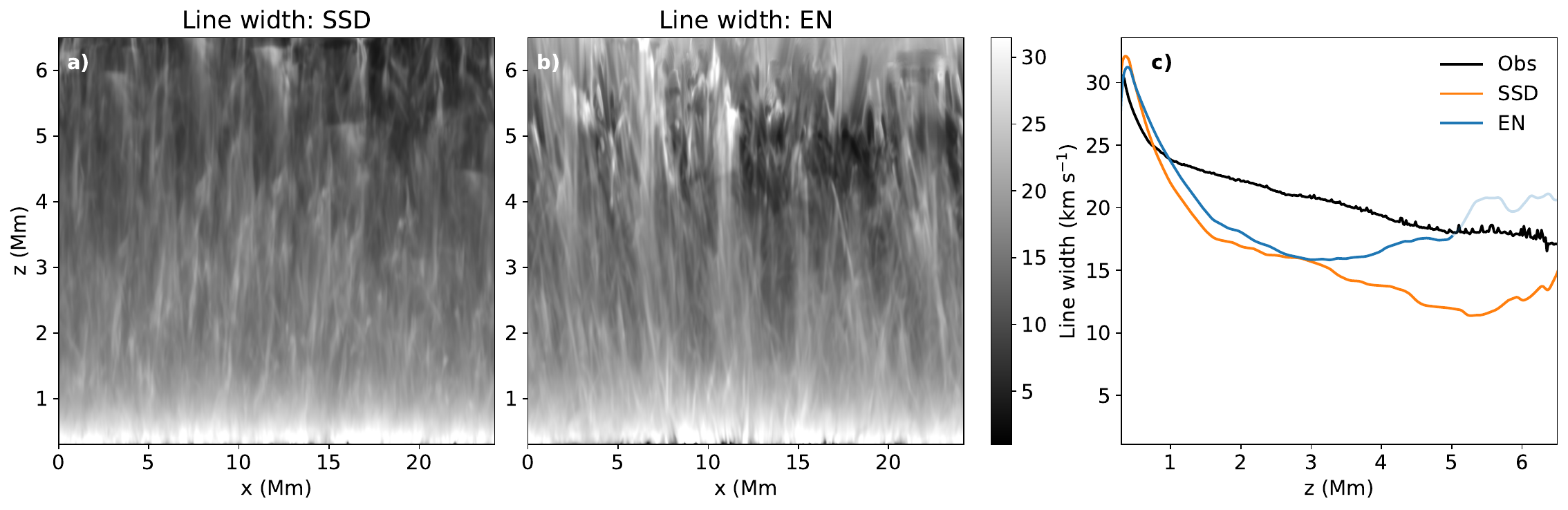}
    \caption{Line width comparison between the synthetic and observed spectra. Panels a) and b): maps of the line width, calculated as the second moment of the line profile, for the SSD and EN simulations, respectively. Panel c) shows the line width variation with height for the time-averaged observed spectra and spatially averaged simulated spectra. The plot for the widht in EN simulation is faded out above 5~Mm to emphasize that line emission there is very weak.}
    \label{fig:width}
\end{figure*}

To compare the observed and synthetic line widths more rigorously, we show, in the panels a) and b) of Fig.~\ref{fig:width}, line width ($\sigma$) maps for the two synthetic spectral cubes. The line width is calculated as the second moment of the line profile and expressed in velocity units. Panel c) shows the variation of $\sigma$ with height for the temporally averaged observed spectrum and the spatially averaged synthetic spectra. The EN simulation shows a larger line width than the SSD one at all the considered heights (note that emission in the EN cube is very weak above 4~Mm heights). We surmise that this difference in widths is the consequence of the plasma flows along the predominantly horizontal magnetic field lines at chromospheric heights. Still, both simulations show widths that are $\approx5$~km/s smaller than the observed. This discrepancy can be caused by our simplified radiative transfer model, or by a lack of horizontal velocity of the dense and cool plasma producing Ca~II emission in the simulated spectra. 

Spicule-like phenomena channel various wave modes into the solar atmosphere. Therefore, the difference in width between the observed and computed line profiles can indicate a discrepancy in the amplitude of the (kink and torsional) waves found in the simulation and in the observed chromosphere. We emphasize that observations like these are a unique diagnostic of horizontal velocities at chromospheric heights. Namely, contrary to feature tracking applied to disk-center observations \citep[see e.g.,][and references therein]{Kostic_2026_linetracking}, observations at the limb directly probe the horizontal velocities via the Doppler effect. At the limb, these horizontal flows could manifest as waves or field-aligned flows, depending on the magnetic field inclination. Specifically, thanks to the very fast cadence of \susi{} ($\approx0.25$~s), we can rule out temporal smearing as a source of broadening (see Fig.~\ref{fig:line_time_evolution}). Therefore, the comparison between these observations and synthetic spectra serves as a unique constraint for comprehensive simulations such as the one considered here. We note that, in principle, the line width discrepancy can be offset by introducing additional microturbulent velocity in the synthetic spectra, such as the one typically found in chromospheric on-disk diagnostics \citep[e.g.][]{da_Silva_Santois_2020_multithermal}. 

Another source of discrepancy can come from the assumption of the constant source function (Eq.~\ref{eq:const_slab}). Optically thick structures along the LOS will result in a source function gradient where the source function increases inside the structure. This would, in turn, increase the emergent intensity in the line wings and thus the perceived width of the lines. Consistent modeling of these effects requires a full 3D radiative transfer approach \citep[][]{Leenaarts_2009_multi3D} to obtain a more physically grounded source function distribution. To our knowledge, such 3D modeling has not been attempted for off-limb observing geometries yet.

\section{Conclusions}\label{sec:conclusion}

This letter presents a set of high-spatial- and high-temporal-resolution observations of the Ca~II~K emission at and above the solar limb. These observations allow us to analyze the spectral structuring of the line emission several Mm above the limb, at a very fast cadence (0.25~s) that was hard to achieve for this wavelength range before the \susi{} instrument \citep[see also][]{CastellanosDuran2026ApJ}. Note that the Interface Region Imaging Spectrograph \citep[IRIS, ][]{IRIS_2014_IRIS} observed Mg~II~h\&k lines above the limb, but at significantly lower spatial and temporal resolution \citep[e.g.,][]{Alissandrakis_2018_IRISspicules, Tei_2020_IRIS}. While there are many similarities between the Mg~II and Ca~II resonant lines, the reported Mg~II emission reaches higher than what \susi{} observed for Ca~II, which can be explained by the higher opacity of the Mg~II lines \citep[see also][ for a direct comparison between IRIS and Hinode/SOT spicule observations]{Pereira_2014_spicules}. The main finding of \susi{} observations is that the emission reaches beyond 6~Mm above the limb and exhibits multiple peaks with substantial velocity shifts between them, pointing to several distinct structures with sufficient Ca~II~K opacity to produce such emission. The line shifts reach velocities of $\approx30-50$~km/s, indicating strong horizontal motions, similar in magnitude to those found in spicules \citep{Pereira_2012_quantifying_spicules, De_pontieu_2012_CaIIHspicules}.

We attempted to model this emission using two distinct radiative transfer approaches. A one-dimensional spherical geometry synthesis using a FAL-C model of \citet{Fontenla_1993_falc} can qualitatively describe the behavior of the line on-disk and just above the limb, but completely lacks the emission above the start of the transition region in the FAL-C model ($\approx$1.5~Mm heights). This is because one-dimensional average models don't have sufficient opacity in the line in the higher layers, and do not account for the temporal variation of the chromosphere. We recall that such models are inherently biased, as indicated already by \citet{Carlsson_1995_chromo}. The second approach uses simplified three-dimensional radiative transfer in a model atmosphere from a 3D rMHD simulation. In this case, dynamics driven by the turbulent magnetoconvection supply cool and dense plasma at the required heights to qualitatively reproduce the extent of the line emission above the limb. This reinforces the importance of treating the chromosphere as a multi-dimensional, inhomogeneous, and dynamic medium. However, we still find a discrepancy in the spectral line width, indicating either the lack of strong enough horizontal velocities in the simulation, or a necessity for more detailed radiative transfer modeling, or both. 

\begin{acknowledgments}

I.M. acknowledges the financial support from the Serbian Ministry of Science and Technology through the grants 451-03-33/2026-03/200104 and 451-03-33/2026-03/200002. L.P.C. gratefully acknowledges funding by the European Union (ERC, ORIGIN, 101039844). C.M.J.O is grateful to the Royal Astronomical Society’s Norman Lockyer Fellowship and the University of Glasgow’s Lord Kelvin/Adam Smith Leadership Fellowship for financially supporting this work. F. A. I. is a member of the “Carrera del Investigador Científico” of CONICET and supported by MPG through the Max Planck Partner Group between MPS and the University of Mendoza, Argentina.

\sunrise{} is supported by funding from the Max-Planck-Förderstiftung (Max Planck Foundation), NASA under Grants \#80NSSC18K0934 and \#80NSSC24M0024 ("Heliophysics Low Cost Access to Space" program), and the ISAS/JAXA Small Mission-of-Opportunity program and JSPS KAKENHI Grant Numbers JP18H05234 and JP23K25916. This research has received financial support from the European Union’s Horizon 2020 research and innovation program under grant agreement No. 824135 (SOLARNET). It has also been funded by the Deutsches Zentrum für Luft- und Raumfahrt e.V. (DLR, grant no. 50 OO 1608). The Spanish contributions have been funded by the Spanish MCIN/AEI/10.13039/501100011033 under projects RTI2018-096886-B-C5, PID2021-125325OB-C5, and PID2024-156066OB-C5, and from "Center of Excellence Severo Ochoa" awards to IAA-CSIC (SEV-2017-0709, CEX2021-001131-S), all co-funded by "ERDF A way of making Europe".
\end{acknowledgments}

\bibliography{references}
\bibliographystyle{aasjournalv7}

\end{document}